# Several AES Variants under VHDL language In FPGA

Sliman Arrag[1], Abdellatif Hamdoun [2], Abderrahim Tragha [3] and Salah eddine Khamlich [4]

[1] Department of Electronics and treatment of information
UNIVERSITE HASSAN II MOHAMMEDIA, Casablanca, Morocco

[2] Department of Electronics and treatment of information
UNIVERSITE HASSAN II MOHAMMEDIA, Casablanca, Morocco

[3] Department of computing and Mathematics
UNIVERSITE HASSAN II MOHAMMEDIA, Casablanca, Morocco

[4] Department of Electronics and treatment of information
UNIVERSITE HASSAN II MOHAMMEDIA, Casablanca, Morocco

**Abstract**
This paper provides four different architectures for encrypting and decrypting 128 bit information via the AES. The encryption algorithm includes the Key Expansion module which generates Key for all iterations on the fly, Double AEStwo-key triple AES, AESX and AES-EXE. These architectures are implemented and studied in Altera Cyclone III and STRATIX Family devices.

## 1. Introduction

Data security becomes an important factor for a wide range of applications, including communication systems, wireless devices, and many other embedded applications [1-3] being the one taken into account by their hardware application and by their integration into the recent communication systems.

Several of the encryption algorithms have been developed [2-4]. Keeping pace with the changing technology security hackers, electronic eavesdropping and violation-emails were coming in the field with new techniques to attack the security devices [17], [19]. And to protect against any attack from the unusual source of useful information and their transmission, the standard encryption algorithm AES (Advanced Encryption standard), standard of a Federal Information Processing Standard is approved by National Institute of Standards and Technology [4], [7], [8], [11]. AES has 10 tower complex algebraic and matrix operations that provide high processing power and a delay in the encryption and decryption. That is why at the beginning of this work, the speed is considered a major problem is given on hardware-based execution. Field Programmable Gate Array based implementation is chosen in this work as a field Programmable gate arrays offer lower cost, flexibility and reasonable income as Application Specific Integrated Circuit implementation. researcher has proposed implementing the AES processor on the safety features of many FPGA files since early version of the FPGA on the market was weak capacity .Now other manufacturer FPGA input level. Recently the design of an AES processor using VHDL and implementation on FPGAs without sacrificing the security feature of the algorithm is reported [6]. Altera FPGA is another known for clients. Literature [10], [12], [13], [18], [20], [23], [24] describe design and implementation of AES processing in the FPGA.

## 2. AES ALGORITHM DESCRIPTION AND ANALYSIS

To comprehend the interior functioning of AES [5], [9] it appears discriminating to start with taking into consideration it in its entire (diagram block).The diverse operations will be retailed sequentially subsequently. On the other side exist are three other architectures AES algorithm (128,192 & 256 bits), in consequence, all examples to come and illustrate the mechanisms will use a key size of 128 bits.

The Advanced Encryption Security algorithm is a symmetric block code. It is distinct for a block of 128 bits and key size of 128, 192 & 256 bits. After to the key size, these numbers of the AES are called AES_128, AES_192 and last AES_256. This article based on the implementation of AES 128, which is most frequently used AES diverse. On the other hand, the existing architecture can also be used for the other key sizes. The succeeding subsection describes the AES transformations, which are the structure blocks of AES encryption and decryption (fig. 1).

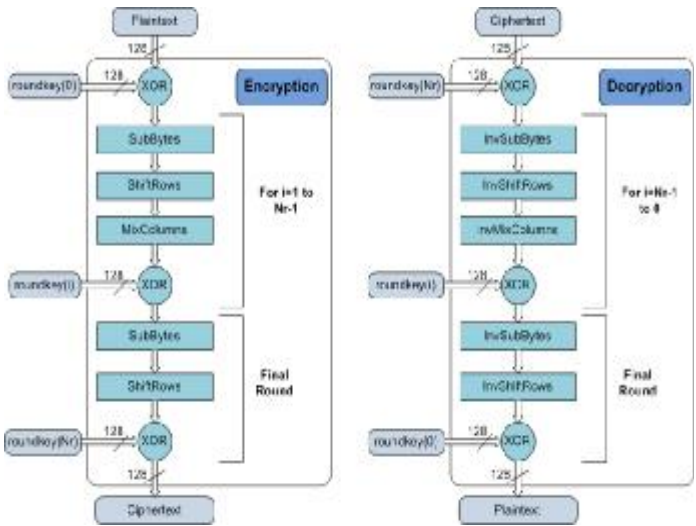

Fig. 1 the Advanced Encryption-Decryption Standard.

Algorithm AES can be cut in 3 blocks:

**Initial __Round:** It is the first and simplest of the stages. It only counts one operation: Add--Round--Key.

Observation--number one: The inverse of this operation bloc is herself.

**N__Rounds:** such as N gives the number of iterations. These numbers vary under the dimension of the key used. (128 bits use N=9 for 192 bits use N=11 and for 256 bits use N =13. This second stage is constituted of N iterations including each the four following operations: Sub--Bytes, Shift--Rows, Mix--Columns, and Add--RoundKey.

**Final__Round:** This phase is nearly the same to one of the N iterations of the second stage. The only variation is that it doesn't consist of the operation Mix Columns.

### 2.1 ADDROUN-key

In this transformation, N key is added to the operational State by a simple bit wise XOR operation (that is a summation in Galois). Each around key consists of four words from the key size schedule procedure.

### 2.2 SubBytes

In This non-linear transformation every byte of the working State using a substitution table (Sbox) [3]. We used this Sbox, which is reversible, by composing two transformations [9]:

• A. Taking the multiplicative opposite in the finite field GF($2^8$) by:

$$m(x) = x^8 + x^4 + x^3 + x + 1 \quad (1)$$

• As irreducible polynomial; the element {00} is mapped on itself.

B. applies an affine (under GF (2) [1], [8]) transformation distinct by:

$$b_i' = b_i \oplus b(i+4)\mod 8 \oplus b(i+5)\mod 8 \oplus b(i+6)\mod 8 \oplus b(i+7)\mod 8 \oplus c_i \quad (2)$$

• For $0 \leq i < 8$, where $b_i$ is the **i**-th bit of the byte and $c_i$ is the **i**-th bit of a invariable byte **c** with the value {63}.

• Observation--number two: The function opposite named Inv_SubBytes consists in applying the identical function but this point in time while using Inv_SBox that is the opposite table of SBox.

### 2.3 ShifRows

In this transformation, the bytes in the finishing three rows of the prepared State are cyclically shifted over diverse numbers of bytes. The first row, row" 0", is not shifted. Row one of the operational State is left shifted by 1 byte place; row two is left shifted by two byte positions; row three is left shifted by three byte positions.

• Observation--number three: The inverse function of this Inv_ShiftRoxs operation consists in replacing the shift on the left on the right by a shift.

### 2.4 MixColumns

This transformation [9] operates on the operational State column-by-column, treating each column as a four-term polynomial over GF($2^8$). These polynomials *are* multiplied with a unchanging polynomial a(x), precise in the standard.

| 02 | 03 | 01 | 01 | D4 | E0 | B8 | 1E |
|----|----|----|----|----|----|----|----|
| 01 | 02 | 03 | 01 | BF | B4 | 41 | 27 |
| 01 | 01 | 02 | 03 | 5D | 52 | 11 | 98 |
| 03 | 01 | 01 | 02 | 30 | AE | F1 | E5 |

Fig. 2 Example of Multiplication the Mixcolumn.

*Result [1,1] = 02•D4 XOR 03•BF XOR 01•5D XOR 01•30*

• Observation--number four: The inverse function of this operation, InvMixColumns, based in multiplying the State_out matrix by the inverse of the stable matrix.

$$\begin{pmatrix} 0E & 0B & 0D & 09 \\ 09 & 0E & 0B & 0D \\ 0D & 09 & 0E & 0B \\ 0B & 0D & 09 & 0E \end{pmatrix}$$

Fig. 3 Matrice of Multiplication inverse the Mixcolumn.

## 3. MULTIPLE ENCRYPTION: SEVERAL AES VARIANTS

In this section, we give the descriptions of several AES variants that are designed such that we can have longer keys. These constructions use multiple encryptions to realize stronger security.

### 3.1 Double AES

Double AES is the simplest variant. In Double AES, we use two keys K1 and K2 and the ciphertext C and plaintext P are computed as follows (Fig. 4):

$$C = EK2 \ (EK1 \ (P))$$
$$P = DK1 \ (DK2(C))$$

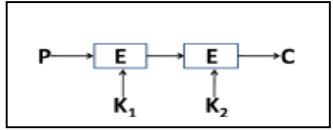

Fig. 4 Double DES.

### 3.2 Two-key Triple AES

In two-key triple AES, we do three times AES operations with two different keys. Note that triple AES is designed such that it becomes equivalent to single AES when K1 is equal to K2 for compatibility. In triple AES, we use two keys K1 and K2 and the ciphertext C and plaintext P are computed as follows (Fig.5):

$$C = EK1 \ (DK2 \ (EK1 \ (P))$$
$$P = DK1 \ (EK2 \ (DK1(C)))$$

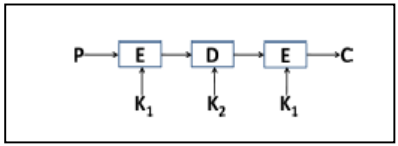

Fig. 5 Triple AES.

### 3.3 AESX

In AESX, we use three keys K1, K2, K3 and the ciphertext C and plaintext P are computed as follows (Fig. 6):

$$C = K3 \ \oplus \ EK2 \ (P \ \oplus \ K1)$$
$$P = K1 \ \oplus \ DK2(C \ \oplus \ K3)$$

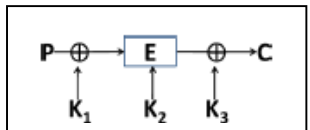

Fig. 6 AESX.

### 3.4 AES-EXE

In AES-EXE, we use three keys K1, K2, K3 and the ciphertext C and plaintext P are computed as follows (Fig. 7):

$$C = EK3 \ (K2 \ \oplus \ EK1 \ (P))$$
$$P = DK1 \ (K2 \ \oplus \ DK3(C))$$

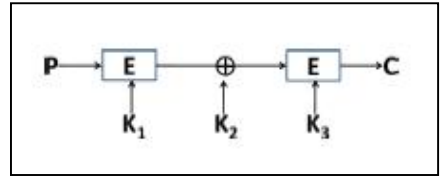

Fig. 7 AES-EXE.

## 4. IMPLEMENTATION OF SEVERAL AES VARIANTS (CIPHERING & DECIPHERING) IN FPGA

Implementation uses the VHDL programming language that currently is frequently a language used very conventional for FPGA [16].

To show that these changes give a better results in conditions of speed and area than the subsequent effort, we evaluate the encryption /decryption codes) based on the Several AES Variants. The comparison considered two criterions: speed and area utilization. The design was implemented on a Cyclone III (EP3C80F780C6 model) device.

The encryption block of aes_128, aesx and double_aes are represented in Figures (8, 9 and 10) where the main signals used by the implementation are shown.

The design and the software of the simulation is Quartus II.

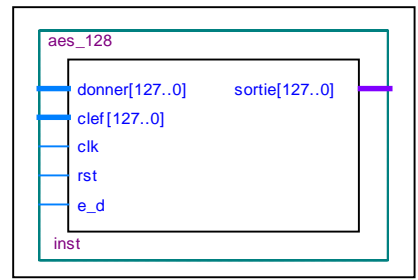

Fig. 8 AES (ciphering & decoding) block.

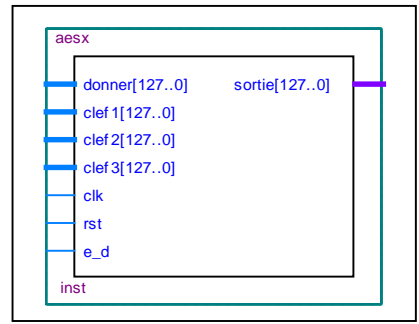

Fig. 9 AESx (ciphering & decoding) block.

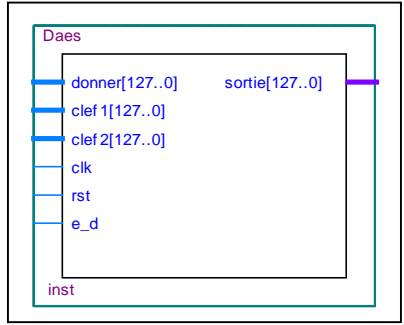

Fig. 10 Double-AES (ciphering & decoding) block.

The main goal of this hardware implementation is not speed, but the area & resource limitations of a specific target FPGA device, respectively Xilinx Stratix (model EP1S80B956C6 ) which is a low resource platform, namely: 79040 Total logic elements, 692 I/O buffers, 12 PLL, 22 DSP,2 DLL, 16 Global Clock.
And also we use Circuit Cyclone III (EP3C80F780C6 model) which is a low resource: 81264 logical elements; 430 pine to in /out; Embedded Multiplier 9-bit elements 488; capacity of memory 2810880bits; 4PLL.
The main signals are: the clock of the system (CLK), the system reset (RST), signal of the load which load the key and given them and signal that permits to encode and to decipher given them. A summary of the occupied resources is presented in the comparative table. (Table 1)

Table 1: Comparative table between different implementation constitute AES algorithm

| Implementation | FPGA Device | | | | |
| --- | --- | --- | --- | --- | --- |
| | Total pins | Total logic elements | Peak virtual memory Megabyte | Total registers | Total memory bits |
| Aes_128_arch1 | 387 | 75840 | 370 | 128 | 0 |
| Aes-128_arch2 | 387 | 80829 | 582 | 128 | 0 |
| Aes_128_arch3 | 387 | 75147 | 368 | 128 | 0 |
| AESx_128 | 643 | 75270 | 401 | 128 | 0 |
| Double_AES_128 | 515 | 149945 | 753 | 128 | 0 |

**Notes:**

• Aes_128_arch1: Mixcolumn architecture the AES-128 based on the methods Mathematical application [14] [22].

• Aes_128_arch2: Mixcolumn architecture the AES-128 based on the methods Galois Multiplication lookup tables [15] [22].

• Aes_128_arch2: Mixcolumn architecture the AES-128 based on the methods Properties of the binary calculation [22].
The implementation uses the VHDL programming language, which nowadays is a well-established commonly used language for FPGAs. The design & simulation software is Xilinx ISE 10.1.

• According to the comparative table we can observe that with first construction of AES-128 of the setting in. implementation occupy more that (75840 slices) of the device, when the second need of the setting in. implementation about (80829 slices) of capacity total of the device, on the other hand the third architecture of AES-128 has need (75147tranches) of the device.
In the fourth implementation AESx_128 (Mixcolumn architecture the AESx_128 based on the methods Properties of the binary calculation) of the setting in. implementation roughly (75270 slices) of capacity total of the device. In fifth implementation double_AES_128 128 (Mix column architecture the double_AES_128 based on the methods Properties of the binary calculation), of the setting in. implementation occupies more that (149945 slices) of the device.
The first conclusion is that the third implementation (Mix column architecture the AES_128_arch3 based on the methods Properties of the binary calculation) bet in implementation is more efficient than architecture of the first and the second, about the number of occupation of resources of the device.
The second conclusion is that the two last implementation when the most reliable security
(AESx_128 we use three key different, Double_AES_128 we use two different encryption key with encryption two times).
We find the firth Architecture bet in implementation is more efficient than architecture of the fifth, about the number of occupation of resources of the device.

## 5. SIMULATION & INTERPRETATION

The diagrams retailed of the simulation the processes for the setting in implementation AES_128 are presented below, in Figure (12 and 13). The length total of the process of the ciphering is (44s) and some decoding is (97s), otherwise simulation of AESx_128, Presented below, in figure (14 and 15).encryption is the time to (61s) and some decoding (114s), and the end Double_AES_128 simulation presented below, in figure (16 and 17). The length total of the process of the ciphering is (121s) and some deciphering is (187s).

Figures 11,12,13,14,15 and 16: shows the simulation results of the encrypt and decrypt data(128 bits) by using three architectures different (Aes_128_architectures 1,2 and 3),than first extension AESx_128 we using three keys different , than last extension double AES_128 when we use two keys different, in all this implementation and simulation we give 1bit e_d (to control the encryption and decryption ,e_d=1 Ciphering else Deciphering ), 128 bit plaintext (data that we need to encrypt) and 128 bit key (for each key used also to generate another keys), of course clock (clk) was used to synchronize the various blocks and rst representing the reset the given if equals 1, finally, we will have in 128 bit output (cyphertext).we use for our simulation QuartusII simulator

• Ciphering :( figure11 )

**e_d**:1
**Plaintext**: hamdoun_&_tragha
**Key:** arragsliman_miti
**cyphertext**:8[139][195]S[189] :[190]P[206][221][153][132][205]bI*

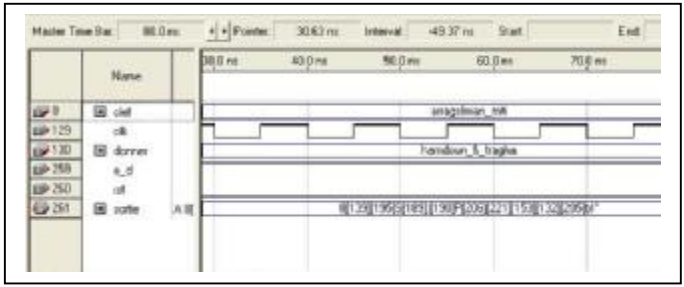

Fig. 11 Simulation of the ciphering of AES-128.

- Deciphering :( figure 12)

**e_d:**0
**plaintext :**8[139][195]S[189] :[190]P[206][221][153][132][205]bI*
**Key:** arraglsiman_miti
**Cyphertext:** hamdoun_&_tragha

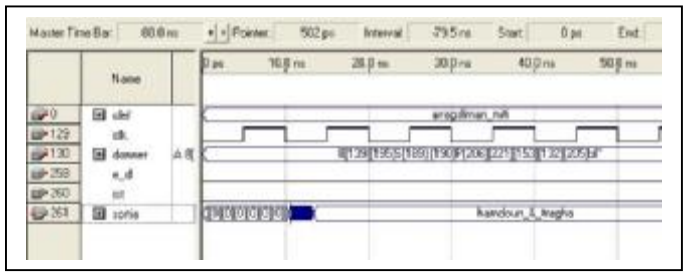

Fig. 12 Simulation of the decoding of AES-128.

- Ciphering :( figure13 )

**e_d**:1
**Plaintext**: hamdoun_&_tragha
**Key1:** arragsliman_miti
**Key2:** Dr__ARRAG_SLIMAN
**Key3:** DR_khamlichsalah
**Cyphertext**: [132][145]A1[22]![203]PM[190]o&siLG

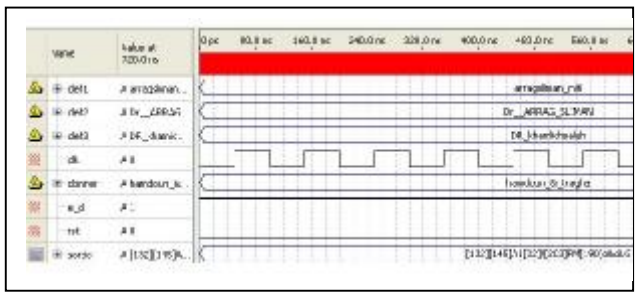

Fig. 13 Simulation of the ciphering of AESx-128.

- Deciphering :( figure 14)

**e_d**:0
**Plaintext**: [132][145]A1[22]![203]PM[190]o&siLG
**Key1:** arragsliman_miti
**Key2:** Dr__ARRAG_SLIMAN
**Key3:** DR_khamlichsalah
**Cyphertext**: hamdoun_&_tragha

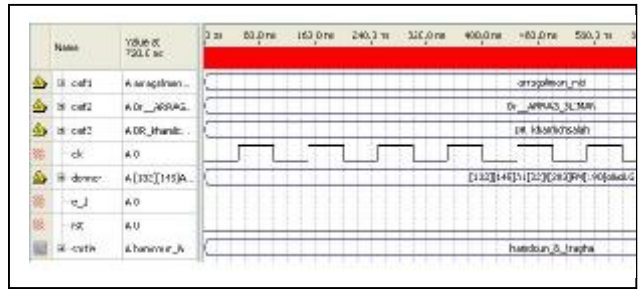

Fig. 14 Simulation of the decoding of AESx-128.

- Ciphering :( figure15 )

**e_d**:1
**Plaintext**: hamdoun_&_tragha
**Key1:** arragsliman_miti
**Key2:** Dr__ARRAG_SLIMAN
**Cyphertext**:[4]f[129][229][224][203][25][154]H[248][211]z([6]&L

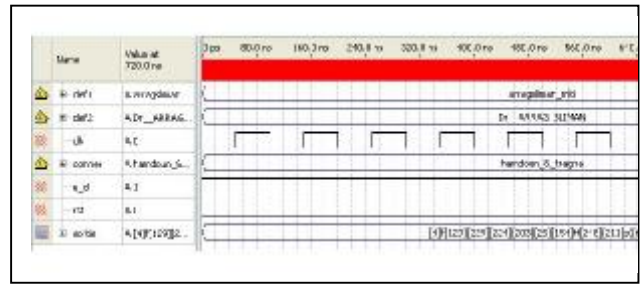

Fig. 15 Simulation of the coding of double_AES_128

- Deciphering :( figure16 )

**e_d**:0
**Plaintext**:[4]f[129][229][224][203][25][154]H[248][211]z([6]&L
**Key1:** arragsliman_miti
**Key2:** Dr__ARRAG_SLIMAN
**Cyphertext** :hamdoun_&_tragha

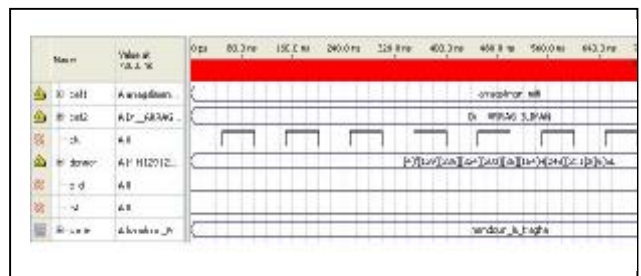

Fig. 16 Simulation of the decoding of double_AES_128

During our implementation, we met more than a few difficulties among which we declare:

- Separation in time of implementation between the ciphering and the decoding for the reason that of the configuration of various functions and we could remedy this difficulty by optimization of the code.

- The troubles of battery overflow caused by the working system, and we consider that it is preferable to work with machines of huge performances (RAM, speed of clock, cache memory.).

- compute it binary in field of Galois notably the multiplication and the inverses of the matrixes.

- The description occasionally dark of the algorithm of RIJNDAEL (AES) and especially in the phase of decoding.

This method of cryptage is used in the domains commercial, industrial and financial and on an ever-increasing number of the PCs. its power on the market increases each day.

Note: during the simulation I used
two different simulators, the first version is Quartus II 9.1 and the second is altered UP Simulator.

## 6. CONCLUSIONS

First conclusion: We showed a construction that we call AESx_128 and it's variant to make double AES secure against the related-key and meet-in-the-middle attacks in practice and the construction is more secure compared with AES_128. Our construction is generic and applicable to any block cipher such as AES to have a longer key effectively. Considering the attacks on AESx_128, we gave the heuristic discussion about the security of the proposed scheme, but further investigation of the security of AESx_128 will be needed in the practical settings.

Second conclusion: our paper has presented a brief description of the FPGA Based Hardware Implementation of the Multiple Encryption algorithm AES, underlining the benefits of this modern design concept. An FPGA implementation of an encryption algorithm is a cryptographic module device in which the structure is software implemented. The FPGA implementations allow us to increase flexibility, lower costs, and reduce time to release enhanced cryptographic equipment, providing a satisfactory level of security for communication applications, or other electronic data transfer processes where security is needed.

## 7. POSSIBLE FUTURE WORK

Possible future work may include more optimizations to the program. Ranking at the top in the CPU SAMPLES profile, the Key Schedule algorithm seems to be taking most of the CPU counts. It's likely that after analyzing the code to this algorithm, the running time of the program could be reduced even further.

Other optimizations might include the tables that are utilized in the rounds for faster encryption. Since all four tables are rotated versions of each other, it might be possible to combine them into just a single table Doing this will enable us to reduce the amount of space needed for the program. It is definitely possible to just use a single table, but that means extra rotation operations in the rounds.